# Symmetry Theory of the Flexomagnetoelectric Effect in the Bloch Lines

**B. M. Tanygin**[a]

[a] Radiophysics Department, Taras Shevchenko Kyiv National University, 4G, Acad. Glushkov Ave., Kyiv, Ukraine, UA-03127

*Corresponding author:* B.M. Tanygin, Radiophysics Department, Taras Shevchenko Kyiv National University, 4G, Acad. Glushkov Ave., Kyiv, Ukraine, UA-03127.

*E-mail*: b.m.tanygin@gmail.com

*Phone*: +380-68-394-05-52

**Abstract.**

It was shown, that there are 48 magnetic point groups of the Bloch lines including 22 (11 time-invariant and 11 time-noninvariant) enantimorphic and 26 non-enantiomorphic groups. The Bloch lines with the time-noninvariant enantiomorphism have identical types (parities) of the magnetization and polarization dependences. The soliton like Bloch lines are derived from the symmetry classification. The tip electrode method of the creation of these Bloch lines is suggested for the potential applications in the magnetoelectric memory devices. The method of the experimental determination of the flexomagnetoelectric properties of the Bloch lines carried by the Bloch domain walls has been suggested. New type of the flexomagnetoelectric coupling, determined by the spatial derivatives of the electric polarization can be found in the vicinity of the Curie temperature or compensation point of the ferrimagnets. The multiple states Bloch line based magnetoelectric/multiferroic memory is proposed. It can be considered as a concept of the magnetoelectric enhancement of existing Bloch line memory invention.

## 1. Introduction

Coupling mechanism between magnetic and electric subsystem in magnetoelectric materials [1] is of considerable interest to fundamentals of condensed matter physics and to the applications in the novel multifunctional devices [2] including low-power-consumption spintronic and magnonic devices. The magnetoelectric coupling is possible between homogeneous magnetization and electric polarization [3-7], and, also, between inhomogeneous magnetization and polarization [8-14]. Latter type of coupling was called as flexomagnetoelectric (FME) interactions [15-17]. The FME coupling is possible not only between existing magnetic and ferroelectric orderings (i.e. in the multiferroic material). It, also, describes inducing of the polarization in the region of the magnetic inhomogeneity [14]. The phenomenological theory of the FME coupling in any symmetry crystals of the cubic, tetragonal and orthorhombic families has been recently developed [18].

FME interaction has been investigated experimentally and theoretically in the micromagnetic structures like spin cycloids [12,19], magnetic domain walls (DWs) [8,9,13,15-17,20], magnetic (anti)vortices [14] and the vertical Bloch lines (preliminary experimental report [21]). The nonvolatile random access memory using a Bloch-line memory element was introduced (e.g. [22]). The FME coupling in the Bloch lines is the additional (electric) mechanism of the random access memory control.

The magnetoelectric effects, particularly, FME ones, are closely related to the magnetic symmetry. The group-theoretical description of the FME coupling in all possible magnetic DWs (including 0° DW [23]) was already suggested in Ref. [24]. The similar domain theory was developed [25]. The investigation of the ferromagnetic [26] and magnetoelectric [24] DW multiplication (degeneration) was the initial investigation towards building of the complete systematization of the Bloch lines. This classification is not yet obtained.

Building of the point symmetry and enantiomorphism classification of the Bloch lines with FME coupling in any magnetic DW (including 0° DW) is the purpose of this work.

## 2. Symmetry classification of the Bloch lines in the magnetoelectric materials

The micromagnetic and ferroelectric structure can be investigated qualitatively without direct solution of the variational problem [13,24]. This concept has been applied recently to all theoretically possible magnetic domain walls (DWs) in the cubic m$\bar{3}$m crystal [24] on the basis of the DWs complete point symmetry classification [26,27]. The influence of the crystal shape and non-local interactions is taking into account automatically using point group of the crystal shape in scope of the group-theory analysis [27]. Important relation between the DW enantiomorphism [28] and type of the FME coupling was recently shown [24]. The same rule is expected for the Bloch lines.

The major feature of the symmetry classification is complete systematization of the DW multiplicity (degeneration) [24,26]. If kinetic [29] (spontaneous) mechanism of the Bloch lines appearing is considered then DW multiplicity schema produces the complete systematization of the magnetic and non-magnetic [29] Bloch lines. Each combination of two equal energy DW segments corresponds to the new Bloch line. Full variety of such segment instances is the symmetry classification essence. Different Bloch line micromagnetic structures can appear for the given pair of the DW segments. Only symmetry approach gives the simplest method of building formal systematization of the all possible Bloch lines.

Since DWs can be considered as thin layers, their symmetry is described by one of the 528 magnetic layer groups [30,31]. To determine the layer's physical properties continuum approximation is used which leads to the point-like layer groups [32]. If continuous translation operation is considered as an identity then the plain DW groups transform to the magnetic point groups $G_k$. There are 64 such groups ($1 \leq k \leq 64$) of the magnetic DWs [26] including 42 groups of 180° DWs [33], 42 groups of the 0° DWs, and 10 groups of DWs with the non-collinear domain magnetizations [26].

Group-theoretical methods permit identification of the DW multiplicity [24,26]: the set of the DWs with identical energies and different structures. Quantitatively, the DW multiplicity at the fixed boundary conditions is determined by the following relation of the magnetic group orders:

$$q'_k = |G_B|/|G_k|, \tag{1}$$

where $G_B$ is a magnetic point group of the magnetic DW boundary conditions, i.e. combination of two neighboring domains, which are spatially separated. According to the theory of the phase transitions, the order parameter occurrence always decreases the medium symmetry into one of the subgroups. According to the Neumann principle, lowest symmetry of the physical properties cannot exceed crystallographic class symmetry. Consequently, the groups' relation is determined by the following expression:

$$G_k \subseteq G_B \subset G_P, \tag{2}$$

where group $G_P$ describes the symmetry of the crystal in the paramagnetic and paraelectric phase [33], which is restricted by the crystal shape symmetry according to Curie principle [27]:

$$G_P = \infty_{[hkl]}/\mathrm{mmm}1' \cap \mathrm{m}\bar{3}\mathrm{m}, \tag{3}$$

where $(hkl)$ is the film or plate crystallographic orientation.

Boundary conditions of the Bloch line are the DW segments with identical energies and symmetries $(G_k)$, but different structures. Here, traditionally, kinetic mechanism of Bloch line appearing is considered [29]. Each DW segment can be combined with $q'_k - 1$ possible instances. So, the total number of the boundary conditions (let we call their symmetry group as $U_B \subseteq G_B$) of the Bloch lines in the given DW is the following:

$$Q_B = (q'_k - 1)q'_k \tag{4}$$

Not all subgroups $U_B \subseteq G_B$ are allowed, because symmetry of the Bloch line depends on the specific DW segments, i.e. depends on the $G_k$ and lost transformations [33], which relates the DW segments.

For the given pair of the DW segments (given boundary conditions of the Bloch line) multiple Bloch lines are possible:

$$Q'_l = |U_B|/|U_l|, \tag{5}$$

where $U_l$ is the $l$-th magnetic point group of the Bloch line. This Bloch lines have identical symmetries and energies but different structures. They are related by the lost transformations $u_i^{(l)}$ (representatives of the coset classes [24,29]) in the series:

$$U_B = u_0^{(l)}U_l + u_1^{(l)}U_l + \cdots + u_{Q_l'}^{(l)}U_l \,, \tag{6}$$

where $u_0^{(l)} \equiv 1$ (identity operation). This structural difference can introduce different types of the order parameter multiplicity in the Bloch lines: ferromagnetic, ferroelectric, ferroelastic and other types [24,29]. The Bloch line can carry one or several such types. Finally, total number of all possible Bloch lines with identical free energies in the given DW is $N_{BL} = Q_B Q_l'$.

The kinetic [29] generation of the Bloch lines at the domain structure appearance is the decreasing or remaining of the magnetic DW boundary conditions symmetry, i.e.:

$$U_l \subseteq U_B \subseteq G_B \subset G_P, \tag{7}$$

Relation between magnetic point group and the allowed functions $\mathbf{M}(\mathbf{r})$ and $\mathbf{P}(\mathbf{r})$ was formulated in Refs. [24,26]. Let us choose the coordinate system XYZ connected with the DW plane: axis Z is the DW plane normal direction; axis Y is directed along the Bloch line. The distribution of the magnetization is $\mathbf{M}(z,x) = M_x(z,x)\mathbf{e_x} + M_y(z,x)\mathbf{e_y} + M_z(z,x)\mathbf{e_z}$. The distribution of the polarization has the same functional dependence. The algorithm of building of the symmetry classification (table 1-3) is the similar as in the Ref. [13,24,26,27]. The different enantiomorphism types [28] are highlighted. Primary, classification describes actual symmetries of the Bloch lines (groups $U_l$). Additionally, it can be considered as a list of all possible groups $U_B$. In the latter case, functional dependence describe relation between the boundary conditions, i.e. limits vectors $\mathbf{M}(x \to \pm\infty)$ and $\mathbf{P}(x \to \pm\infty)$, which are spatially separated. The type of the Bloch lines multiplicity is determined using given classification, similarly as determined for the DW multiplicity [24]. To obtain the type of the multiplicity, it is necessary to compare groups $U_l$ and $U_B$. If magnetization components are different in case of $U_l$ ($U_l \subseteq U_B$) and $U_B$ then the multiplicity is ferromagnetic. If *at least one* polarization component has different types for these groups then multiplicity is ferroelectric. These criteria can be formulated in the alternative way. If $\mathbf{M}(\mathbf{r})$ and/or $\mathbf{P}(\mathbf{r})$ is not the invariant of at least one lost symmetry transformation $u_i^{(l)}$ of the series (6) then multiplicity is ferromagnetic and/or ferroelectric respectively. The ferroelastic details are not described in scope of the present work. The problem of the enantiomorphic properties of the multiplicity is described in the Ref. [28] on the basis of the novel Barron chirality and enantiomorphism definition.

There are 48 magnetic point groups of the Bloch lines including 22 (11 time-invariant and 11 time-noninvariant) enantimorphic and 26 non-enantiomorphic groups. This is the complete point symmetry classification of all possible Bloch lines (ferromagnetic, ferroelectric, ferroelastic and hybrid coupled types) in the planar magnetic DWs of any magnetically ordered medium. The subset of this classification was obtained early for the ferromagnetic Bloch lines in the 180° DWs only [29]. The present results agree with Ref. [29].

The time-noninvariant enantiomorphic Bloch lines can have the same zero components of the magnetization $M_y(z,x)$ and polarization $P_y(z,x)$. The non-enantiomorphic Bloch lines can, also, have zero components of magnetization $M_x(z,x) = M_z(z,x) = 0$ and the polarization $P_y(z,x) = 0$ at the same time (table 3). The Bloch lines with the time-noninvariant enantiomorphism (table 2) have identical types of the $\mathbf{M}(z,x)$ and $\mathbf{P}(z,x)$ dependences.

### 3. FME coupling in the Bloch lines

The present systematization of the Bloch lines describes the FME coupling in any symmetry crystal [29]. Corresponding phenomenological theory was recently reported [18]. In order to verify the symmetry based results, let us consider most high-symmetric crystal. Higher symmetry suppress most FME invariants, which are possible in lower symmetry crystals [18], making number of the phenomenology constants lowest. If the symmetry based prediction is correct in case of the highest-symmetric cubic m$\bar{3}$m crystal then it will be proper in the any symmetry crystals, which have fewer restrictions on the FME free energy invariants.

The polarization induced by the FME effect in the cubic m$\bar{3}$m crystal is described by the following four-constant expression [34]:

$$\mathbf{P} = \chi_e[\tilde{\gamma}_1 \mathbf{e}_i \nabla_i(M_i^2) + \tilde{\gamma}_2 \nabla(\mathbf{M}^2) + \tilde{\gamma}_3(\mathbf{M}\nabla)\mathbf{M} + \tilde{\gamma}_4 \mathbf{M}(\nabla\mathbf{M})]$$ (8)

For the regular micromagnetic problem (constant magnetization magnitude) the second term equals zero. In the framework of the model of the planar magnetic DW with Bloch line, the expression (8) takes the form:

$$P_x = \chi_e\left[\left(\tilde{\gamma}_1 + \frac{\tilde{\gamma}_3 + \tilde{\gamma}_4}{2}\right)\frac{\partial M_x^2}{\partial x} + \tilde{\gamma}_3 M_z\frac{\partial M_x}{\partial z} + \tilde{\gamma}_4 M_x\frac{\partial M_z}{\partial z}\right]$$ (9a)

$$P_y = \chi_e\left[\tilde{\gamma}_3\left(M_x\frac{\partial M_y}{\partial x} + M_z\frac{\partial M_y}{\partial z}\right) + \tilde{\gamma}_4 M_y\left(\frac{\partial M_x}{\partial x} + \frac{\partial M_z}{\partial z}\right)\right]$$ (9b)

$$P_z = \chi_e\left[\left(\tilde{\gamma}_1 + \frac{\tilde{\gamma}_3 + \tilde{\gamma}_4}{2}\right)\frac{\partial M_z^2}{\partial z} + \tilde{\gamma}_3 M_x\frac{\partial M_z}{\partial x} + \tilde{\gamma}_4 M_z\frac{\partial M_x}{\partial x}\right]$$ (9c)

Simple analysis of odd and even functions [24] shows that expressions (9a-c) give the results which correspond to the symmetry based predictions (table 1-3). This statement is correct for 40 magnetic point groups of the Bloch lines with at least two nonzero magnetization components. Remained 8 non-enantiomorphic groups (italicized rows in table 3) describe symmetry of Bloch lines with single magnetization component (in the 0° DWs as a rule) and two polarization components, which appear near the point of the phase transition or compensation point [33]. The full conformity between phenomenology and symmetry based predictions appears if we take into account additional types of the inhomogeneous magnetoelectric interactions with spatial derivatives of the polarization [35]. Thus, experimental investigations of the magnetoelectric effects in the Bloch lines near the point of the phase transition are of interest.

The similar experimental setup [36,37], which was designed for the investigation of the DW shift, can be used for the detection of the corresponding Bloch line shift. It can be the Bloch lines of type $2'_x 2_y 2'_z (2'_y 2_x 2'_z)$ in the Bloch DW, where $M_z(z,x) \neq 0$ only in the Bloch line volume. The mentioned experiments were targeted to the Néel DW, while Bloch DW was considered as structures with lack of the FME features [24,34,37]. So, the proposed here experimental investigation is free from the DW related FME coupling, i.e. it can be used for clear determination of the FME properties of the Bloch lines independently on such properties of the magnetic

DW. The preliminary experiments, which demonstrate electric field induced displacement of the Bloch line along the DW in the garnet films were already shortly reported [21].

All lost operations are members of the paramagnetic phase point group: $u_i^{(l)} \subset G_P$. Applying of such operation to the media does not change the total free energy. If some external influence exists then it is changed by the operation $u_i^{(l)}$ too. Any magnetic point symmetry operation applied to all bodies of system does not change its free energy independently on the physical nature of the interactions. It is a consequence of the isotropic space and reversal time symmetry properties. The latter criteria is correct in the thermodynamically equilibrium states, which are considered in the present investigation of the multiplicity switching. If we defined "state" as some dissipation process then the present theory cannot be directly applied. However, the Bloch line states do not require involvement of such processes into consideration. Thus, the total internal and external interaction free energy is not changed by $u_i^{(l)}$ as well. If external influence is the invariant of at least one operation $u_i^{(l)}$ then multiple states of Bloch line, which are related by $u_i^{(l)}$, cannot be switched by such external influence. If the external influence is not the invariant of *all* lost operations $u_i^{(l)}$ in the series (6) then it breaks the state multiplication making lower free energy for the certain state. Such influence (e.g. homogeneous / inhomogeneous field) is able to produce switching between $Q_l'$ states of the Bloch line at the given boundary conditions. The DW fragments retain unchanged. Consequently, this approach can be utilized to make Bloch-line based memory with $Q_l'$ states. It can be considered as the enhancement of the existing invention [22]. If the DW fragments can be changed as well then total number of states can be increased to the $N_{BL} = Q_B Q_l'$. The complete investigation of the $Q_B$ values in the cubic crystal is given in the Ref. [26].

Similar symmetry arguments regarding controlling of the multiple states of the Néel DW was formulated and compared successfully with experimental results [24]. This DW has the ferroelectric and ferromagnetic multiplicity at the same time, which was controlled using the inhomogeneous electric (different DW shifts) and homogeneous in-plane magnetic field (multiple states switching). Similar approaches and rules exist in case of the Bloch line multiplication. The group-subgroup description of the micromagnetic state and its boundary conditions is general approach. If multiplicity is ferromagnetic, ferroelectric and ferroelastic at the same time then different ways of the corresponding memory bit state switching are possible.

The 0° DWs are the soliton solutions. Predictably, the magnetic memory elements based on the 0° DW were suggested [38]. Complete symmetry systematization was already suggested for these DWs [26]. The ferromagnetic [26] and magnetoelectric DW multiplication [24] have been suggested as well. Magnetic vortices and skyrmions are well-known soliton solutions. These structures have the similar dimensionality [39] as Bloch lines. However, regular Bloch line has conceptual difference. It has different DW segments as boundary conditions. Thus, most Bloch lines are not the soliton entity. The soliton-like Bloch lines divide identical segments of the magnetic DW. Boundary conditions of this Bloch line are the two parts of the single magnetic DW. Such Bloch line can be derived from the symmetry systematization by the following criterion:

$$U_B = G_k \qquad (10)$$

The soliton like Bloch line can be created in the regular DW volume using the tip electrode with electric potential via to the FME coupling. After switching-out the potential, this Bloch line can leave stabilized by the Dzyaloshinskii-Moriya interaction. The latter approach was already investigated [40] on the symmetry basis for similar chiral soliton structures (skyrmions). The Dzyaloshinskii-Moriya interaction has the chiral nature and is possible in the non-centrosymmetric crystals [40]. Consequently, the clear enantiomorphism establishment for all possible Bloch lines (table 1-3) is important for such applications.

The whole discussed theory can be generalized to the weak ferromagnetics with non-collinear magnetic ordering as well as to the antiferromagnetics using algorithms specified in [41] and [33] respectively. The same FME interactions are possible in case of the antiferromagnetic ordering as it was shown for the multiferroic bismuth ferrite $BiFeO_3$ [16]. Also, present symmetry classification predicts structural changes caused by the Bloch line [29] and DW [33] motion. The buckling symmetry investigations of the type [42] can be performed on the basis of the present theory as well.

## 4. Conclusions

Thus, the magnetic point groups allow determining kind of the distributions of the electrical polarization in the Bloch line volume produced by the flexomagnetoelectric effect. There are 48 magnetic point groups of the Bloch lines including 22 (11 time-invariant and 11 time-noninvariant) enantimorphic and 26 non-enantiomorphic groups. The time-noninvariant enantimorphic Bloch lines have identical types of spatial distribution of the magnetization and polarization. There are coincidence between the symmetry predictions and results obtaining from the known term of the FME coupling for most groups.

The multiple states Bloch line based magnetoelectric/multiferroic memory can be considered as a concept of the magnetoelectric enhancement of existing Bloch line memory invention.

### Acknowledgements

I would like to express my sincere gratitude to Acad. V.G. Bar'yakhtar, who inspired me to the present work. I thank Prof. V.F. Kovalenko, Prof. V. A. L'vov, Dr. S. V. Olszewski, and Dr. O. V. Tychko for their helpful discussion and suggestions.

**Table 1.** Types of spatial distribution of electric polarization induced by the flexomagnetoelectric effect in the Bloch lines with the time-invariant enantiomorphism*.

| Magnetic point group | $M_x(z,x)$ | $M_y(z,x)$ | $M_z(z,x)$ | $P_x(z,x)$ | $P_y(z,x)$ | $P_z(z,x)$ |
|---|---|---|---|---|---|---|
| $2'_x 2_y 2'_z$ | (A,S)/A | (S,S)/S | (s,a)/a | (s,a)/a | (a,a)/s | (a,s)/a |
| $2'_y 2_x 2'_z$ | (S,S)/S | (A,S)/A | (a,a)/s | (s,a)/a | (a,a)/s | (a,s)/a |
| $2'_z$ | (F,S)/F | (F,S)/F | (f,a)/f | (f,a)/f | (f,a)/f | (f,s)/f |
| $2'_x$ | (A,F)/F | (S,F)/F | (S,F)/F | (s,f)/f | (a,f)/f | (a,f)/f |
| $2'_y$ | (F,F)/S | (F,F)/A | (F,F)/S | (f,f)/a | (f,f)/s | (f,f)/a |
| $2_y$ | (F,F)/A | (F,F)/S | (F,F)/A | (f,f)/a | (f,f)/s | (f,f)/a |
| $2_x$ | (S,F)/F | (A,F)/F | (A,F)/F | (s,f)/f | (a,f)/f | (a,f)/f |
| $1$ | (F,F)/F | (F,F)/F | (F,F)/F | (f,f)/f | (f,f)/f | (f,f)/f |
| $2_z$ | (f,a)/f | (f,a)/f | (F,S)/F | (f,a)/f | (f,a)/f | (f,s)/f |
| $2_x 2_y 2_z$ | (s,a)/a | (a,a)/s | (A,S)/A | (s,a)/a | (a,a)/s | (a,s)/a |
| $2_z 2'_x 2'_y$ | (a,a)/s | (s,a)/a | (S,S)/S | (s,a)/a | (a,a)/s | (a,s)/a |

\* $(T_z, T_x)/T_{zx}$ defines the even ("S", i.e. symmetric dependence) or odd ("A", i.e. antisymmetric dependence) function, or sum of odd and even functions (F) for the following parity types: $z \rightarrow -z$ (type $T_z$), $x \rightarrow -x$ (type $T_x$) and the simultaneous changes $z \rightarrow -z$, $x \rightarrow -x$ (type $T_{zx}$). The lower-case means that corresponding component equals zero in the magnetic domain volume.

**Table 2.** Types of spatial distribution of electric polarization induced by the flexomagnetoelectric effect in the Bloch lines with the time-noninvariant enantiomorphism

| Magnetic point group | $M_x(z,x)$ | $M_y(z,x)$ | $M_z(z,x)$ | $P_x(z,x)$ | $P_y(z,x)$ | $P_z(z,x)$ |
|---|---|---|---|---|---|---|
| $m_x'$ | (f,a)/f | (F,S)/F | (F,S)/F | (f,a)/f | (f,s)/f | (f,s)/f |
| $m_y'$ | (F,F)/F | (0) | (F,F)/F | (f,f)/f | (0) | (f,f)/f |
| $2_y/m_y'$ | (F,F)/A | (0) | (F,F)/A | (f,f)/a | (0) | (f,f)/a |
| $2_x/m_x'$ | (s,a)/a | (A,S)/A | (A,S)/A | (s,a)/a | (a,s)/a | (a,s)/a |
| $\bar{1}'$ | (F,F)/A | (F,F)/A | (F,F)/A | (f,f)/a | (f,f)/a | (f,f)/a |
| $m_x'm_z'2_y$ | (s,a)/a | (S,S)/S | (A,S)/A | (s,a)/a | (s,s)/s | (a,s)/a |
| $m_y'm_z'2_x$ | (S,F)/F | (0) | (A,F)/F | (s,f)/f | (0) | (a,f)/f |
| $m_z'$ | (S,F)/F | (S,F)/F | (A,F)/F | (s,f)/f | (s,f)/f | (a,f)/f |
| $2_z/m_z'$ | (s,a)/a | (s,a)/a | (A,S)/A | (s,a)/a | (s,a)/a | (a,s)/a |
| $m_x'm_y'2_z$ | (f,a)/f | (0) | (F,S)/F | (f,a)/f | (0) | (f,s)/f |
| $m_x'm_y'm_z'$ | (s,a)/a | (0) | (A,S)/A | (s,a)/a | (0) | (a,s)/a |

**Table 3.** Types of spatial distribution of electric polarization induced by the flexomagnetoelectric effect in the non-enantiomorphic Bloch lines.

| Magnetic point group | $M_x(z,x)$ | $M_y(z,x)$ | $M_z(z,x)$ | $P_x(z,x)$ | $P_y(z,x)$ | $P_z(z,x)$ |
|---|---|---|---|---|---|---|
| $m_x m_z m'_y$ | (A,S)/A | (0) | (s,a)/a | (s,a)/a | (0) | (a,s)/a |
| $m_y m_z m'_x$ | *(0)* | *(A,S)/A* | *(0)* | *(s,a)/a* | *(0)* | *(a,s)/a* |
| $m'_y m_x 2'_z$ | (F,S)/F | (0) | (f,a)/f | (f,a)/f | (0) | (f,s)/f |
| $m'_x m_y 2'_z$ | *(0)* | *(F,S)/F* | *(0)* | *(f,a)/f* | *(0)* | *(f,s)/f* |
| $m_x m_z 2_y$ | (A,S)/A | (a,a)/s | (s,a)/a | (s,a)/a | (s,s)/s | (a,s)/a |
| $m_y m_z 2_x$ | *(0)* | *(A,F)/F* | *(0)* | *(s,f)/f* | *(0)* | *(a,f)/f* |
| $2'_x/m_x$ | (A,S)/A | (s,a)/a | (s,a)/a | (s,a)/a | (a,s)/a | (a,s)/a |
| $2'_y/m_y$ | *(0)* | *(F,F)/A* | *(0)* | *(f,f)/a* | *(0)* | *(f,f)/a* |
| $2'_z/m_z$ | (A,S)/A | (A,S)/A | (s,a)/a | (s,a)/a | (s,a)/a | (a,s)/a |
| $m_y$ | *(0)* | *(F,F)/F* | *(0)* | *(f,f)/f* | *(0)* | *(f,f)/f* |
| $m_x$ | (F,S)/F | (f,a)/f | (f,a)/f | (f,a)/f | (f,s)/f | (f,s)/f |
| $m_z m'_y 2'_x$ | (A,F)/F | (0) | (S,F)/F | (s,f)/f | (0) | (a,f)/f |
| $m_z m'_x 2'_y$ | (a,a)/s | (A,S)/A | (S,S)/S | (s,a)/a | (s,s)/s | (a,s)/a |
| $m_z$ | (A,F)/F | (A,F)/F | (S,F)/F | (s,f)/f | (s,f)/f | (a,f)/f |
| $m_y m'_x m'_z$ | *(0)* | *(S,S)/S* | *(0)* | *(s,a)/a* | *(0)* | *(a,s)/a* |
| $m_x m'_y m'_z$ | (S,S)/S | (0) | (a,a)/s | (s,a)/a | (0) | (a,s)/a |
| $m_y m'_z 2'_x$ | *(0)* | *(S,F)/F* | *(0)* | *(s,f)/f* | *(0)* | *(a,f)/f* |
| $m_x m'_z 2'_y$ | (S,S)/S | (s,a)/a | (a,a)/s | (s,a)/a | (s,s)/s | (a,s)/a |
| $2_y/m_y$ | *(0)* | *(F,F)/S* | *(0)* | *(f,f)/a* | *(0)* | *(f,f)/a* |
| $2_x/m_x$ | (S,S)/S | (a,a)/s | (a,a)/s | (s,a)/a | (a,s)/a | (a,s)/a |
| $2'_z/m'_z$ | (S,S)/S | (S,S)/S | (a,a)/s | (s,a)/a | (s,a)/a | (a,s)/a |
| $2'_x/m'_x$ | (a,a)/s | (S,S)/S | (S,S)/S | (s,a)/a | (a,s)/a | (a,s)/a |
| $2'_y/m'_y$ | (F,F)/S | (0) | (F,F)/S | (f,f)/a | (0) | (f,f)/a |
| $\bar{1}$ | (F,F)/S | (F,F)/S | (F,F)/S | (f,f)/a | (f,f)/a | (f,f)/a |
| $2_z/m_z$ | (a,a)/s | (a,a)/s | (S,S)/S | (s,a)/a | (s,a)/a | (a,s)/a |
| $m_z m'_x m'_y$ | (a,a)/s | (0) | (S,S)/S | (s,a)/a | (0) | (a,s)/a |